\documentclass[final,3p,twocolumn]{elsarticle}

\usepackage{amssymb}
\usepackage[normalem]{ulem}
\usepackage{graphicx}
\usepackage{xcolor}
\usepackage{xspace}
\usepackage{bm}
\usepackage{gensymb} 
\usepackage{textcomp} 
\usepackage{comment} 

\begin{document}

\begin{frontmatter}

\title{The nontrivial effects of annealing on superconducting properties of Nb single crystals}

\author[Ames,ISUPhysics]{Amlan Datta}
\author[Ames]{Kamal R. Joshi}
\author[Temple]{Giulia Berti}
\author[Ames,ISUPhysics]{Sunil Ghimire}
\author[ISUPhysics]{Aidan Goerdt}
\author[Ames,ISUPhysics]{Makariy A. Tanatar}
\author[Ames]{Deborah L. Schlagel}
\author[Ames]{Matthew F. Besser}
\author[Ames]{Dapeng Jing}
\author[Ames,ISUPhysics]{Matthew Kramer}
\author[Temple]{Maria Iavarone}
\author[Ames,ISUPhysics]{Ruslan Prozorov\corref{cor1}}
\ead{prozorov@ameslab.gov}
\cortext[cor1]{Corresponding author}

\affiliation[Ames]{organization={Division of Materials Engineering, Ames National Laboratory},
            city={Ames},
            postcode={50011}, 
            state={IA},
            country={U.S.A.}}

\affiliation[ISUPhysics]{organization={Department of Physics \& Astronomy, Iowa State University},%
            city={Ames},
            postcode={50011}, 
            state={IA},
            country={U.S.A.}}

\affiliation[Temple]{organization={Department of Physics, Temple University},
            city={Philadelphia},
            postcode={19122}, 
            state={PA},
            country={U.S.A.}}

\begin{abstract}
The effect of annealing on the superconducting properties of niobium single crystals cut from the same master boule was studied by local and global magnetic measurements, as well as scanning tunneling microscopy (STM). The formation of large hydride precipitates was observed in unannealed samples. The variation in structural and magnetic properties was studied after annealing under high vacuum at 800\,\celsius\: for 3 hours, 1400\,\celsius\: for 3 hours, and near the melting point of niobium (2477\,\celsius) for a few seconds. The initial samples had a high hydrogen content. Low-temperature polarized optics and magneto-optical studies show that the formation of large niobium hydride precipitates is suppressed already by 800\,\celsius+3 h annealing. However, the overall superconducting properties in the annealed samples did not improve after annealing, and in fact, worsened. In particular, the superconducting transition temperature decreased, the upper critical field increased, and the pinning strength increased. Parallel studies were conducted using STM, where the sample was annealed initially at 400\,\celsius, measured, annealed again at 1700\,\celsius, and measured again. These studies revealed a ``dirty'' superconducting gap with a significant spatial variation of tunneling conductance after annealing at 400\,\celsius. The clean gap was recovered after annealing at 1700\,\celsius. It is likely that these results are due to oxygen redistribution near the surface, which is always covered by oxide layers in as-grown crystals.

Overall, the results indicate that vacuum annealing at least up to 1400\,\celsius, while expected to remove a large amount of hydrogen, introduces additional nanosized defects, perhaps hydride precipitates, that act as efficient pair-breaking and pinning centers. The dimensionless scattering rate is estimated to have increased from $\Gamma=0.2$ to about $\Gamma=0.4$ after annealing at 1400\,\celsius. These results on single crystals differ drastically from those from polycrystalline bulk niobium (i.e. cut from superconducting radio-frequency (SRF) cavities), where annealing has significant positive effects that are attributed to the improvement of crystalline structure masking the more subtle influence of the hydrides. 
\end{abstract}

\end{frontmatter}


\section{Introduction}

Macroscopic quantum coherence makes superconductors highly appealing for quantum information science (QIS) \cite{Grezes2016,Kjaergaard2020,Huang2020,He2021,Siddiqi2021,Xiong2022,Ezratty2023} and accelerator-related technologies \cite{padamsee2009,Chao2014,Anne_2022}. Niobium is a frequently used superconductor due to its high thermal conductivity at low temperatures in the normal state, the highest superconducting transition temperature $T_{c}\approx9.35\:\mathrm{K}$ among elemental metals, and the ability to carry microwaves with minimal losses in its superconducting state \cite{Stromberg1965,Finnemore1966,Daams1980,Bahte1998,Koethe2000,Prozorov2006a,Kozhevnikov2017,Liarte2017}. These properties make niobium a good choice for Josephson junction-based qubits in thin-film two-dimensional (2D) architecture \cite{wendin2017quantum} and superconducting radio-frequency (SRF) cavities traditionally used in particle accelerators \cite{Padamsee2008,Gurevich2012,Singer2014,Gurevich2017a,alex20,Ueki2022,Ueki2022a}, but more recently in quantum informatics applications in cavity quantum electrodynamics (cQED) modes \cite{Blais2004,Paik2011,Reagor2016}. Since its discovery in the 1930s \cite{Daunt1937}, elemental niobium has become one of the most studied superconductors \cite{Finnemore1966,Kneisel2015}. However, there are still fundamental aspects of niobium properties that necessitate further investigation, both experimental and theoretical. Recent examples are first-principles calculations of its anisotropy in normal and superconducting states \cite{Zarea2023}, and derived from this work the suggestion that, intrinsically, the clean limit niobium is a type-I superconductor \cite{Prozorov2022}.

Like all refractory metals, niobium presents some physical-chemical challenges that can complicate or even obstruct its use in applications. For example, one of the most critical issues in Nb SRF cavities is the so-called ``hydrogen $Q-$disease'' - a significant reduction in the quality factor $Q$, apparently due to niobium hydrides \cite{Knobloch2003,Barkov2012,Barkov2013}. Niobium has a significant affinity for hydrogen and can be charged up to five atomic percent \cite{Buck1971,Schober1975,Isagawa1980,Isagawa1980a,Dzyuba2014}. Hydrogen may come from various hydrogen-rich sources, for example, during liquid-assisted cutting and polishing, and even from ambient moisture. Since all applications related to the superconducting state of niobium require cooling to low temperatures, it is important to understand the formation of various phases, in particular, niobium hydrides and their effect on the superconducting properties \cite{Schober1975,Vinnikov1982,Barkov2012,Barkov2013,Koszegi2017}. In addition, hydrogen is a possible source for two-level systems (TLS), e.g., quantum systems that can exist with the quantum superposition of two independent states, which are the main sources of quantum decoherence in QIS-related devices and structures. The study of TLS-related losses represents a significant portion of contemporary research in applied superconductivity \cite{Burnett2016,Mueller2019,McRae20}.

A traditional approach to the removal of contaminants from metallic solids is heat treatment. Despite a very large body of work, in the case of niobium, there is still significant uncertainty in understanding the evolution of properties with annealing performed at different temperatures and following different protocols. Moreover, device-oriented studies were mostly performed on polycrystalline samples of bulk niobium cut from SRF cavities or whole cavities. In these cases, significant variations in superconducting properties are dominated by re-crystallization, grain growth, fusion, inter-diffusion and other major morphological and structural changes \cite{Tedmon1965,Halbritter1988,Antoine2019,Koethe2000,DangwalPandey2021}.  Single crystals are a good baseline for studying the more subtle effects of hydrogen and other contaminants. Indeed, this aspect was already well understood in the 1960s \cite{Tedmon1965,Halbritter1988}. The novelty of the present work lies in the experimental methods used.

In an unannealed sample with a high load of hydrogen, niobium hydrides of various shapes form below 190\:K as very large precipitates, up to hundreds of micrometers in size, easily visible in an optical microscope \cite{Vinnikov1982,Barkov2013,Koszegi2017}. If the hydrogen load is not large, round or square precipitates of sub-micrometer size are formed, leading to local plastic deformations and nucleation of interstitial prismatic dislocation loops \cite{Schober1975}. 

It was reported that after annealing at 800\:\celsius\: in vacuum for a few hours, a significant amount of hydrogen was removed and it was assumed that niobium hydrides no longer pose a problem \cite{Isagawa1980,Hakovirta2001,Barkov2012,Singer2014}. However, in this paper, we show that other superconducting properties indicate a reduced performance after annealing at intermediate temperatures (compared to the melting point), likely because of a continuing detrimental influence of residual hydrogen in the system. Diffusion of oxygen from the surface to the bulk is also possible. However, it is unlikely that a few nanometer thick surface oxide layers can provide enough oxygen to affect the bulk properties of thick crystals. A simple estimate shows that if all oxygen from, say, 5\:nm thick oxide layers covering a $1 \times 1 \times 1\:\textrm{mm}^3$ diffuses in, it will produce a volume atomic percentage of less than $1\times 10^{-4}$\,at.$\%$, which is negligible. We also note that many earlier studies used annealing near and above 2000\,\celsius\: as a starting point to obtain high-quality samples with a high residual resistivity ratio, $R(300\,\textrm{K})/R(T_c)>100$ \cite{Finnemore1966, Buck1971,Schober1975,Ooi2021}. On the other hand, it is important to look for approaches that are as simple as possible, but still produce acceptable results. This is especially important for potential applications and mass production of devices. For example, annealing at 2300\,\celsius\: requires quite specialized thermal and vacuum equipment and is much more difficult to carry out compared to annealing at 800\,\celsius.

In this paper, we study the effects of annealing at 800\,\celsius, 1400\,\celsius, 1700\,\celsius\: and near the melting point of Nb (2477\,\celsius) on the superconducting properties of single crystals of niobium, all cut from the same single crystal boule. The results suggest that each potential application should develop its own optimized annealing protocol. For example, contrary to common sense, large hydrides are not too detrimental to the overall superconducting properties of the surrounding material but certainly impact the total surface impedance, crucial for SRF cavities. We find that annealing at intermediate temperatures leads to a reduction in superconducting performance manifested in suppression of the transition temperature, $T_{c}$, an inhomogeneous "dirty" superconducting gap with in-gap states, an increase in the upper critical field, $H_{c2}$, and elevated pinning. Therefore, our findings have important implications for the development of a wide spectrum of niobium-based superconducting devices, both for accelerators and for quantum informatics applications.

\section{Material and methods}

\subsection{Niobium single crystals}

Niobium single crystals were cut from a large single crystal boule (approximately $12\times12\times50\:\mathrm{mm^{3}}$) grown at the \textit{Materials Preparation Center} at the \textit{Ames National Laboratory} in 1989. This original single crystal boule was grown from a high purity polycrystalline Nb using an arc-zone melting technique with the resultant orientation of the long axis within 2\degree of {[}310{]} direction \cite{Lograsso1991}. Laue diffraction study indicated a high degree of crystallinity of the sample. The sample morphology was further studied using high-resolution electron microscopy \cite{King1990}.

The master crystal has been stored on a shelf since then. Multiple smaller samples were cut from the boule over the years using different methods. The original cuts used electrical discharge machining (EDM) with a paraffinic hydrocarbon as the cooling fluid. The samples for our study were cut using a diamond wire saw with Rock-Oil as lubricant. The final polishing / etching steps included vibratory polishing in a suspension of alumina; electropolishing in a dilute aqueous solution of 5\% aqua regia; acid etching in a solution of hydrochloric acid mixed with hydrogen peroxide. At least some of these etchants (i.e. hydrochloric acid) remove niobium oxide, enabling access to the fresh metal surface. Therefore, as a result of handling, storage and many hydrogen-containing chemicals involved, the initial, ''as-grown" single crystals had a high level of hydrogen, which was readily identified when large formations of niobium hydride were observed at the first cooling below 190\:K. 

The samples studied were shaped as flat cuboids with the two largest dimensions of the order of 3-5\:mm and a thickness of 1-2\:mm. For this work, the orientation of the crystals was not important, although we know that we had {[}100{]} and {[}110{]} directions perpendicular to flat faces. 

\begin{figure}[tb]
\centering
\includegraphics[width=7cm]{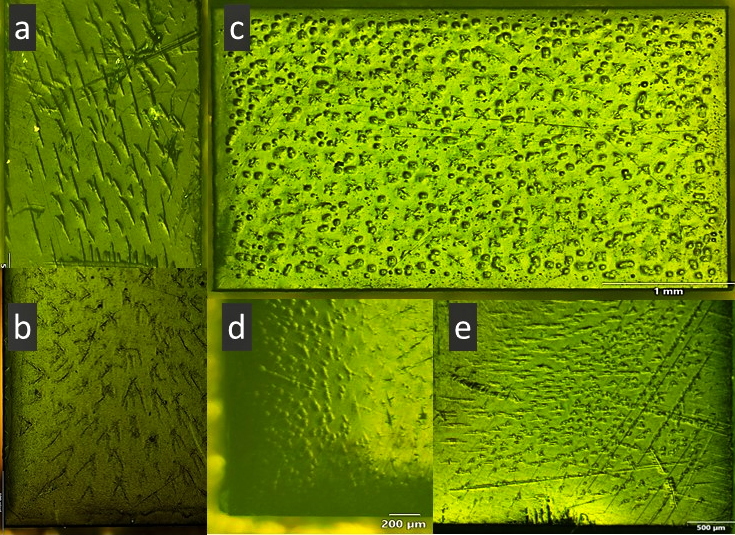} 
\caption{Large hydride precipitates in niobium single crystals observed in a linearly polarized light microscope. (a) After cooling to $T=4\:\mathrm{K}$ and (b) when the same sample was warmed up to room temperature showing the scars left by large hydrides formed upon first cooling. (c,d,e) Images of different crystals at $T=4\:\mathrm{K}$. No systematic correlation between the crystallographic orientation and the shape of hydrides is found. (a), (b) and (c) are {[}100{]} oriented perpendicular to the page; (d) and (e) are {[}110{]} oriented. }
\label{fig1:POhydrides}
\end{figure}

It is crucial to control and record the first cooling. If a sample was cooled for measurements prior to annealing, large niobium hydride precipitates formed below 190\:K. Some were up to 100\,{\micro m} in size and had different shapes: cubes, four-point stars, boomerangs, chicken feet, worms, and droplets. Figure\,\ref{fig1:POhydrides} shows polarized light images of the variety of morphologies observed. Panel (a) shows the $100 - 150\,\micro\mathrm{m}$ long surface formations. Apparently, they are so large that they irreversibly deform the crystal lattice. When warmed up, the remnants (which we call ``scars'', other papers ``skeletons'' \cite{Schober1975}) of this significant plastic deformation are clearly visible, as shown in Fig.\ref{fig1:POhydrides}(b). We found that these scars on the surface cannot be removed by annealing at any temperature and require re-polishing or re-melting to be removed from the surface. Figure\,\ref{fig1:POhydrides}(c,d,e) shows polarized light images of different crystals at 4\:K. No systematic direct correlation was found between the crystallographic orientation and the shape of the hydrides. Figure\,\ref{fig1:POhydrides} (a), (b) and (c) shows samples that are {[}100{]} oriented perpendicular to the page; (d) and (e) are {[}110{]} oriented. The shape of the hydride precipitates is probably determined by some residual strain from the crystal growth and depends on where the particular crystal was located inside the master boule.

\subsection{Heat treatment}

Different types of annealing protocols were used for the comparative study using optical and magnetization methods. Two protocols involved vacuum annealing at 800\,\celsius\: and 1400\,\celsius, respectively.  A separate protocol was used for STM studies whose main purpose was to remove the surface oxide layer. One of the STM samples was partially melted in an ultra-high vacuum (UHV) chamber, with a base pressure of $1\times10^{-11}$\,torr, during an attempt to remove surface oxygen and was then subjected to optical and magnetic measurements. 

One Nb single crystal was first measured at low temperatures and then annealed at 800\,\celsius\: for three hours, and measured again. The crystal was spot-welded onto a tantalum sample plate using tantalum wires and then loaded into an Omicron ultra-high vacuum (UHV) chamber for cleaning and annealing. The base pressure of the UHV chamber is $2\times10^{-11}$\,torr. Before vacuum annealing, the sample was sputtered with a 1 kV monoatomic Ar$^{+}$ beam to remove surface contamination and the native oxide layer. After removal of carbon and oxygen, confirmed by X-ray photoelectron spectroscopy, the sample was annealed in UHV at 800\,\celsius\: for 3 h. Heating to 800\,\celsius\: at a rate of 6\,\celsius/min took roughly 2 hours; cooling was faster, at 10\,\celsius/min. 

Four other crystals were annealed simultaneously in one run. The crystals were tightly wrapped in tantalum foil and put together in an alumina crucible. The chamber was evacuated using a diffusion pump down to $1\times10^{-11}\,\textrm{torr}$. The heating and cooling ramp rates were 700\,\celsius\: per hour. The furnace was heated to 1400\,\celsius\: and held there for 3 hours. The next day, after the furnace completely cooled overnight, still actively maintaining a high vacuum, the chamber was back-filled with dry air for passivation to facilitate the formation of the surface oxide layer for protection from moisture. For the planned "global'' measurements that obtain the signal from sample volume, this a few nanometers thick layer did not contribute. Samples were quickly moved to the glovebox, unwrapped, and placed in their containers. Each sample was exposed to air for only a few minutes maximum before measurements. The following four samples were made. \uline{Sample\:A}: $5.56\times2.91\times0.90\,\mathrm{mm^{3}}$, {[}110{]} normal to the flat face. This sample was not cooled prior to vacuum annealing at 1400\,\celsius\: for 3 hours. No hydrides were observed in optical microscopy when the sample was cooled after annealing. 
\uline{Sample\:B}: $5.56\times3.44\times0.90\:\mathrm{mm^{3}}$, {[}110{]} normal to the flat face. (Step 1) Cooled to 4\:K and measured before any annealing. Polarized light imaging, as well as magneto-optical imaging, revealed large hydrides that appeared during the first cooldown at 190\:K. (Step 2) The sample was annealed at 1400\,\celsius. When cooled again, no large hydrides were observed in optical microscopy. However, the scars left by the hydrides formed in Step 1 were clearly visible.  
\uline{Sample\:C}: $3.12\times3.00\times1.09\:\mathrm{mm^{3}}$, {[}111{]} normal to the flat face. (Step 1) Similar to sample B, cooled first before any annealing. Low temperature measurements revealed large hydrides; (Step 2) Annealing at 800\,\celsius\: for 3 hours. After this annealing, visible hydrides did not appear, but their remnants (scars) remained. (Step 3) The sample was annealed at 1400\,\celsius\: together with samples A and B. Again, no hydrides were observed optically. The scars from the hydrides formed in Step 1 were still visible and did not change visually from Step 2.  
\uline{Sample\:D}: $4.00\times2.50\times1.50\:\mathrm{mm^{3}}$, {[}100{]} normal to the flat face. (Step 1) Cooled to 4\:K and measured before annealing. Low-temperature imaging revealed large hydrides. (Step 2) {[}accidental{]} annealing in UHV at 2477\,\celsius\: for several seconds during the attempt to remove surface oxygen for tunneling studies. Part of the sample melted, but part remained solid. We were able to cut out and measure this part. Visible hydrides were not observed after cooling. However, the appearance of scars left by large hydrides from Step 1 was practically unchanged, indicating that only a complete re-melting can remove those plastic deformations of the crystal lattice. 

The annealing process for the samples studied with STM was conducted within a UHV chamber directly attached to the STM chamber. This setup facilitated the transfer of samples to the STM without the need to break the vacuum. Using this method, the sample surface was preserved, effectively preventing contaminations that could affect STM measurements.  The treatment here aimed to remove any adsorbed material and, in particular, the surface oxide layer. A Nb (100) single crystal was mounted on an e-beam sample holder made of tantalum and cleaned with sputtering and annealing cycles in an ultra-high vacuum (UHV) system with a base pressure of low $1\times10^{-11}\,\textrm{torr}$. Inside the UHV system, the sample was first sputtered for approximately 2 hours at room temperature at 1.5\,keV under a partial argon pressure of $1\times10^{-5}\,\textrm{torr}$. The sample was then slowly heated for several hours until the maximum temperature of 400\,\celsius\: was reached at which the sample was kept for a maximum of 10 minutes. The temperature of the sample was measured using an optical pyrometer set outside the UHV chamber. After this treatment, the sample was transferred to the attached STM chamber for STM investigation. 

After STM measurements, the sample was transported again to the room temperature UHV chamber, without breaking the vacuum, for another sputtering cycle and several annealing cycles up to 1700\,\celsius. During each flash annealing, the pressure in the UHV chamber was kept below $8\times10^{-8}\,\textrm{torr}$.

\subsection{Experimental Methods }

\subsubsection{Low-temperature polarized optics and magneto-optics}

Optical studies were performed in a low-temperature polarized light setup consisting of the closed-cycle optical cryo-station Model-s50 from \emph{Montana Instruments} and the \emph{Olympus BX3M} optical system with long focal length objectives. The sample was placed on top of a gold-plated cold finger inside the vacuum chamber and could be directly observed through two optical windows. This completely closed-cycle system allows controlled measurements from 3.8\:K and up to room temperature. Optical characterization was performed in two modes. (1) Direct study of the sample surface in a polarized light microscope with nearly crossed polarizer and analyzer axes, which greatly enhanced the contrast of surface features. We refer to this measurement simply as "polarized optics'' (PO). In particular, large hydride formations are clearly observable, as is shown in Fig.\ref{fig1:POhydrides}. (2) The second mode is based on the magneto-optical Faraday effect, and we call it a "magneto-optical'' (MO) study. Here, we use specially fabricated magneto-optical indicators, multilayered structures with the active ferrimagnetic bismuth-doped iron-garnet layer with in-plane magnetization. A gold or aluminum mirror layer is deposited at the bottom to reflect incident light back to the objective. The indicator is placed on top of the sample studied. Linearly polarized light propagates through the transparent magnetic layer, reflects from the mirror, and returns to the objective. As a result, it acquires a double (light travels the thickness twice) Faraday rotation by an angle proportional to the component of magnetization along the light path. This $z-$component of magnetization appears only in response to the $z-$component of the magnetic induction on the sample surface. Therefore, a two-dimensional picture revealed by this double Faraday effect in the garnet maps the distribution of the $z-$component of the garnet magnetization, which in turn mimics the distribution of the $z-$component of the magnetic induction on the sample surface. In all images, brighter corresponds to higher magnetic induction. Black is a complete screening, $B=0$. More details can be found elsewhere\cite{Prozorov2006a}.

\subsubsection{Magnetic measurements}

Magnetization was measured using a \emph{Quantum Design} vibrating sample magnetometer (VSM) in a 9 T \emph{Quantum Design} Physical Property Measurement System (PPMS). The instrument allows true sweeping of the external magnetic field at a controlled rate, which is very useful for studies of flux dynamics in superconductors. To minimize the differences in the out-of-plane demagnetizing factors, all samples were measured with a magnetic field applied along the extended surface. The total measured magnetic moment, $m$, was normalized, $M=\gamma m$, to obtain the magnetization $M$ in gauss. Here, the calibration constant $\gamma=4\pi\left(1-N\right)/V$, where $V$ is the volume of the sample in cm$^{3}$. The demagnetizing factor, $N$, of a cuboidal sample with dimensions $a\times b\times c$, along the $c-$axis, is given by, $N_{c}\approx\left(1+3c\left(a^{-1}+b^{-1}\right)/4\right)^{-1}$\cite{Prozorov2018}. Comparing the calibration factor $\gamma$ with the slope of the initial magnetization at low temperatures, $-d\left(4\pi m\right)/dH$, we found excellent numerical agreement for all samples.

\subsubsection{Scanning tunneling microscopy (STM)}

STM measurements were performed in a Unisoku STM system. Pt-Ir tips were used in all experiments. They were tested on Au (111) films before imaging the Nb single crystal. Spectroscopic measurements of differential conductance were performed with a lock-in technique with an ac modulation of 0.2$\:$mV and a frequency $f=373.1\:\mathrm{Hz}$ superimposed on the sample bias at $T=1.5\:\mathrm{K}$. 

\section{Results}

\subsection{Polarized Optical and Magneto-optical imaging}

First, we show the variety of different morphologies of niobium hydrides in single crystals cooled before any annealing. Polarized light images of the sample surface are presented in Fig.\ref{fig1:POhydrides}. The (false) green color is due to the green-light monochromatic filter used for all our magneto-optical imaging. The natural appearance of the niobium crystals is metallic-gray.

\begin{figure}[tb] 
\centering
\includegraphics[width=7cm]{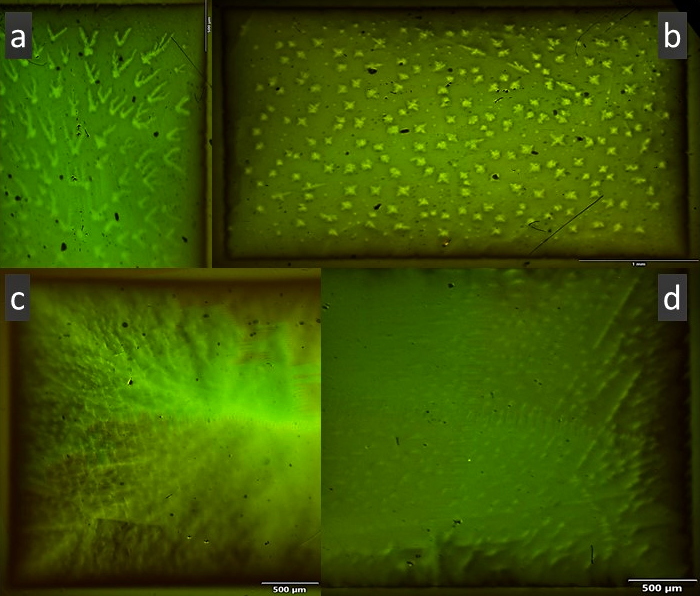} 
\caption{Magneto-optical images of the trapped (remnant) magnetic flux in Nb single crystals. (a,b,d) Samples were cooled in a 400\:Oe magnetic field to 4\:K, then the field was turned off and the image recorded. (c) The same protocol but cooling in a 320\:Oe field to 8\:K then the field was turned off. Panels (a,b) are {[}100{]} oriented, (c,d) are {[}110{]} oriented. (We emphasize that sample surface is not visible in magneto-optical images, since there is a mirror sputtered at the bottom of the magneto-optical indicator.)}
\label{fig2:MOhydrides}
\end{figure}

Figure \ref{fig2:MOhydrides} shows the magneto-optical image of the remnant (trapped) magnetic flux. These are the same crystals as those used to compose the image in Fig.\ref{fig1:POhydrides}. In the experiment, a few hundred oersted magnetic fields was applied above the superconducting transition temperature, the sample was cooled down, and the magnetic field was turned off. Due to omnipresent pinning, the superconductor retains a large amount of magnetic flux in the form of Abrikosov vortices. The images in frames (a), (b) and (d) show trapped flux when a magnetic field was turned off at 4\:K, whereas in image (c) the magnetic field was turned off at 8\:K. Frames (a), (b), and (d) show a fairly uniform background flux distribution, indicative of low pinning. The large hydrides are clearly visible. They are (likely) not superconducting, but they appear bright because they trap magnetic flux, as would any cavity surrounded by the supercurrents. Frame (c) shows a greater variation of the flux density, indicating greater inhomogeneity at temperatures approaching $T_c$. (We emphasize that the sample surface is not visible in magneto-optical images, since there is a mirror sputtered at the bottom of the magneto-optical indicator.)

\begin{figure}[tb] 
\centering
\includegraphics[width=7cm]{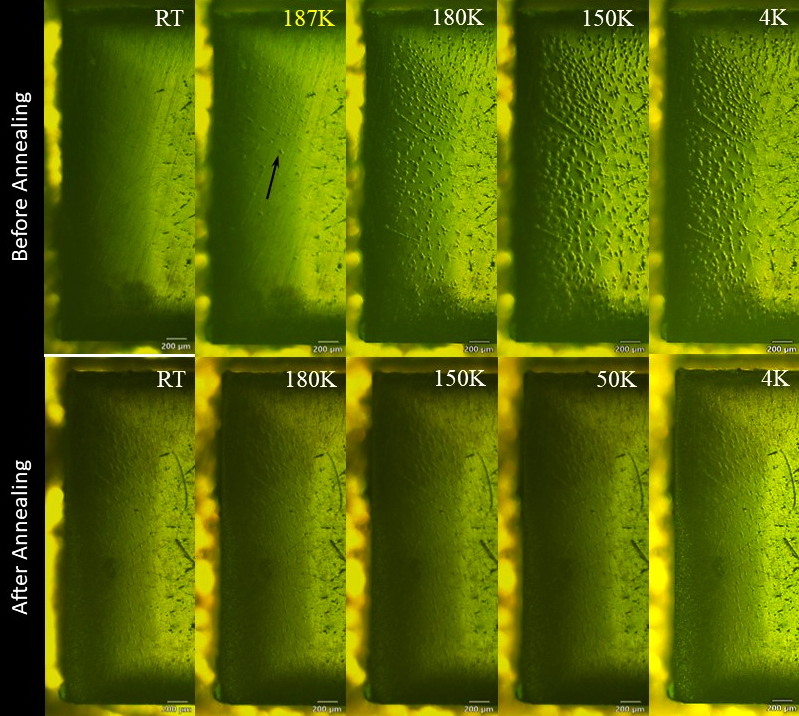} 
\caption{Polarized light imaging of sample B at different temperatures. Top row: appearance of niobium hydrides at around 187\:K, their quick growth and stabilization already at 150\:K. An arrow points to the first hydride we could resolve visually at 187\:K. Bottom row: similar imaging but after annealing at 1400\,\celsius. No visible hydrides are formed,
but the scars are visible at all temperatures.}
\label{fig3:MOb4after}
\end{figure}

Now we examine the process of nucleation of large hydrides upon cooling. Figure \ref{fig3:MOb4after} shows the series of polarized light images taken at different temperatures in sample B that was cooled before subsequent annealing at 1400\,\celsius. The upper row of Fig.\ref{fig3:MOb4after} shows that the first detectable hydrides appear around 187\:K. Hydrides grow fast in size and stabilize at 150:K, without any changes thereafter. The bottom row shows the same sample B after it was annealed at 1400\,\celsius. At all temperatures, some scars from the hydrides can be seen, but no actual formation of new hydrides is observed.

\begin{figure}[tb] 
\centering
\includegraphics[width=7cm]{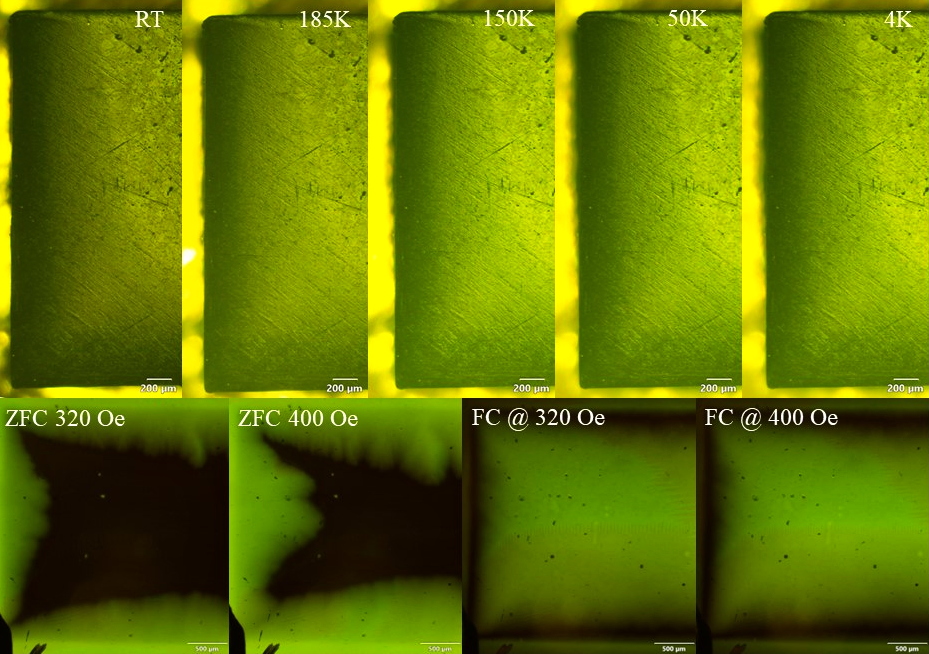} 
\caption{Imaging of sample A annealed at 1400\,\celsius\: before first cooling. Top row: no hydrides are visible in the polarized light optical images at any temperature upon cooldown. Bottom row: no hydrides are observed in magneto optical imaging. Two left frames of the bottom row show magnetic flux penetration when 320\:Oe and 400\:Oe were applied after zero-field cooling to 6\:K, respectively. Two right frames show remanent trapped magnetic flux after the sample was first cooled to 4\:K in the indicated magnetic field, then the magnetic field was turned off and the image recorded.}
\label{fig4:POcooling}
\end{figure}

We now examine sample A, which was first annealed at 1400\,\celsius\: and only then was it cooled for the first time. The result is shown in Fig.\ref{fig4:POcooling}. The upper row of Fig.\ref{fig4:POcooling} shows the sequence of polarized light images recorded at different temperatures that show no visible change and therefore no signature of the formation of large hydrides. The bottom row of Fig.\ref{fig4:POcooling} shows magneto-optical imaging in the superconducting state. The first two images were obtained after cooling the sample in a zero magnetic field to 4\:K and applying the indicated magnetic field to study flux penetration. This is known as the zero-field-cooled (ZFC) process. The last two images were taken in the same magnetic field, but the sample was first cooled in the indicated magnetic field to 4\:K, and then this field was removed to reveal the trapped magnetic flux. This is known as the field-cooled (FC) remanent state. These two types of magneto-optical imaging are used to estimate the magnetic flux gradient, hence the shielding persistent current density, which allows us to discuss the pinning strength in the material.

\begin{figure}[tb] 
\centering
\includegraphics[width=7cm]{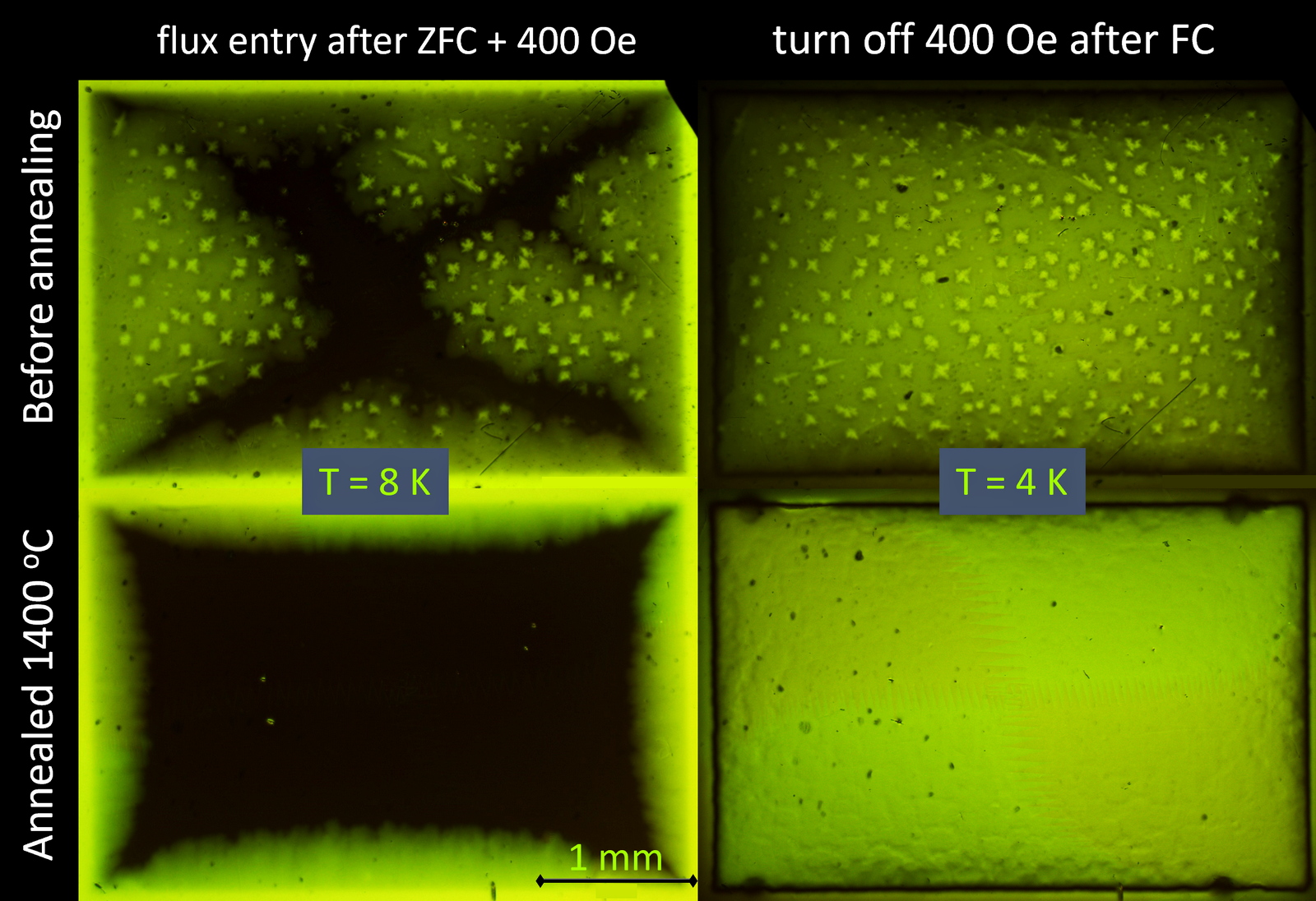} 
\caption{Magneto-optical images of a sample before (top row) and after (bottom row) the annealing at 1400\,\celsius. The first column shows magnetic flux penetration after zero-field cooling to 8\:K whence a 400\:Oe magnetic field was applied. The second column shows the remanent trapped magnetic flux after cooling to 4\:K in a 400\:Oe magnetic field and turning the field off.}
\label{fig5:POMOannealed}
\end{figure}

Finally, Fig.\ref{fig5:POMOannealed} compares flux penetration and flux trapping before (top row) and after (bottom row) annealing. The left column shows the penetration of magnetic flux in a magnetic field of 400\:Oe applied after cooling in a zero field to 8\:K. The large hydride precipitates are clearly visible in a sample prior to annealing. It appears that these hydride formations do not hinder flux penetrations, which is natural considering their large size being irrelevant for the pinning of vortices. The flux front reaches much further into the sample before annealing at 1400\,\celsius, top left image of Fig.\ref{fig5:POMOannealed}. The flux penetrates to a much smaller depth after annealing, which indicates a stronger pinning. This is unexpected after annealing, which usually reduces the pinning strength. The second column of Fig.\ref{fig5:POMOannealed} shows trapped magnetic flux after the sample was cooled in a 400\:Oe magnetic field to 4\:K and the field was turned off. Again, the hydrides in the sample before annealing do not seem to play any role in determining the flux distribution. The image after annealing shows a large amount of trapped flux, again consistent with elevated pinning. 

Note that spatially resolved vortex pinning on hydrides was studied by Bitter decoration forty years ago \cite{Vinnikov1982}. Interestingly, in that study the researchers could not determine whether the hydrides were normal or superconducting. They found that vortices enter the hydride precipitates in integers of a magnetic flux quantum. Of course, even in a normal cavity embedded in a superconductor bulk, the magnetic flux is quantized. Bulk niobium hydrides are not superconducting. However, small hydrogen-rich precipitates could become superconducting by the superconducting proximity effect with the niobium matrix. This question deserves further microscopic study, perhaps using STM and other scanning magnetic probes. 

In summary, polarized optics and magneto-optical studies of single crystals before and after annealing show that already 800\,\celsius\ treatment prevents the formation of large visible niobium hydrides. However, this cannot be taken as evidence that all hydrogen was removed by this treatment. The magneto-optical images reveal that contrary to expectations, vortex pinning not only did not decrease but actually increased following the annealing process. This means that large hydrides do not interact magnetically with the surrounding niobium matrix. Annealing surely removes a large percentage of hydrogen, leading to the disappearance of these large hydride formations. However, some dilute uniformly distributed background hydrogen remains, and its density might even increase after annealing, leading to the formation of efficient pinning centers with sizes that match the coherence length. To investigate this hypothesis, we turn to magnetic measurements.

\subsection{Magnetization}

\begin{figure}[tb] 
\centering
\includegraphics[width=7cm]{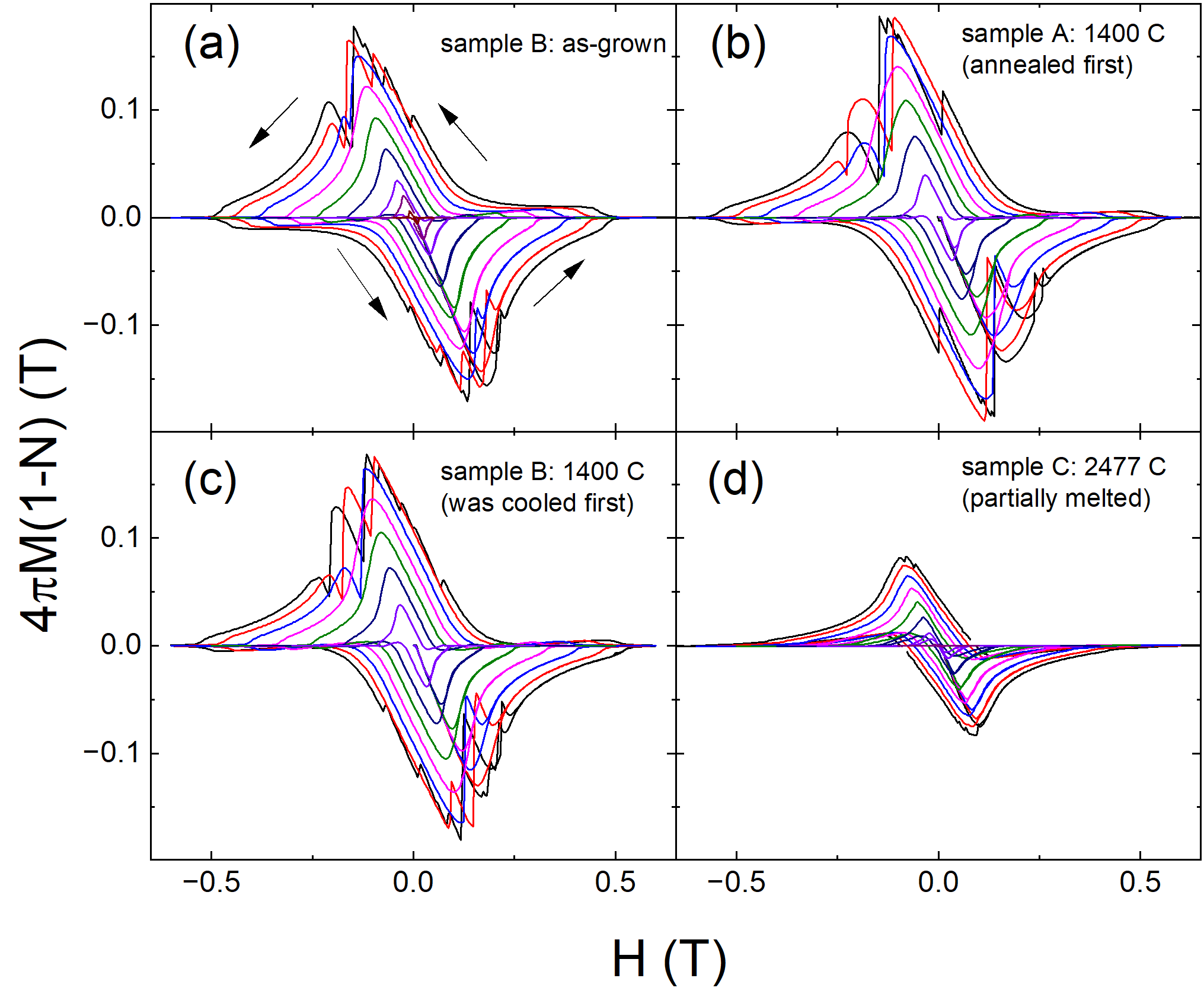} 
\caption{Magnetization loops measured at a 1\:K interval between 2 and 9\:K in samples A, B and C. In all frames, vertical and horizontal scales are the same, which allows for a visual comparison of the magnitude of magnetization before and after annealing. Evidently, only partial melting forced a substantial change.}
\label{fig6:AllTsLoops}
\end{figure}

Magnetization loops recorded at temperatures from 2\:K to 9\:K with a 1\:K interval are shown in Fig.\ref{fig6:AllTsLoops}. To enable visual comparison, the loops are plotted using the same vertical and horizontal scales. The largest effect is observed in the partially melted sample, where the irreversible part of magnetization (the difference between the descending and ascending branches) is drastically reduced, indicating a significant decrease in the pinning strength. In contrast to this expected behavior, in annealed samples, the pinning did not decrease and even increased slightly after annealing, which defies common sense. Especially since previous studies of polycrystalline niobium showed a significant reduction in pinning strength after annealing \cite{Dzyuba2014,Isagawa1980}. Naturally, it was interpreted to be due to significant morphological changes during annealing leading to a more perfect crystal structure, unrelated to the pining of vortices on impurities, which dominates our measurements. 

A pronounced pinning-related feature of Fig.\ref{fig6:AllTsLoops}(a,b,c) is the observation of thermomagnetic instabilities \cite{Young2005,Prozorov2006a}. These sudden magnetic flux jumps occur when the critical current density is large and heat dissipation from fast-moving vortices has nowhere to dissipate, resulting in a catastrophic thermomagnetic runaway. This may even lead to a total collapse of the critical state, so large that jumps in sample temperature could be measured \cite{Prozorov2006a}. Unfortunately for superconducting applications, niobium is prone to thermomagnetic instabilities more than any other superconductor. This instability is observed for all sample geometries and degree of crystallinity, including polycrystalline bulk, thin and thick foils, and thin films \cite{Young2005,Prozorov2006a,Joshi2022a}.

\begin{figure}[tb] 
\centering
\includegraphics[width=7cm]{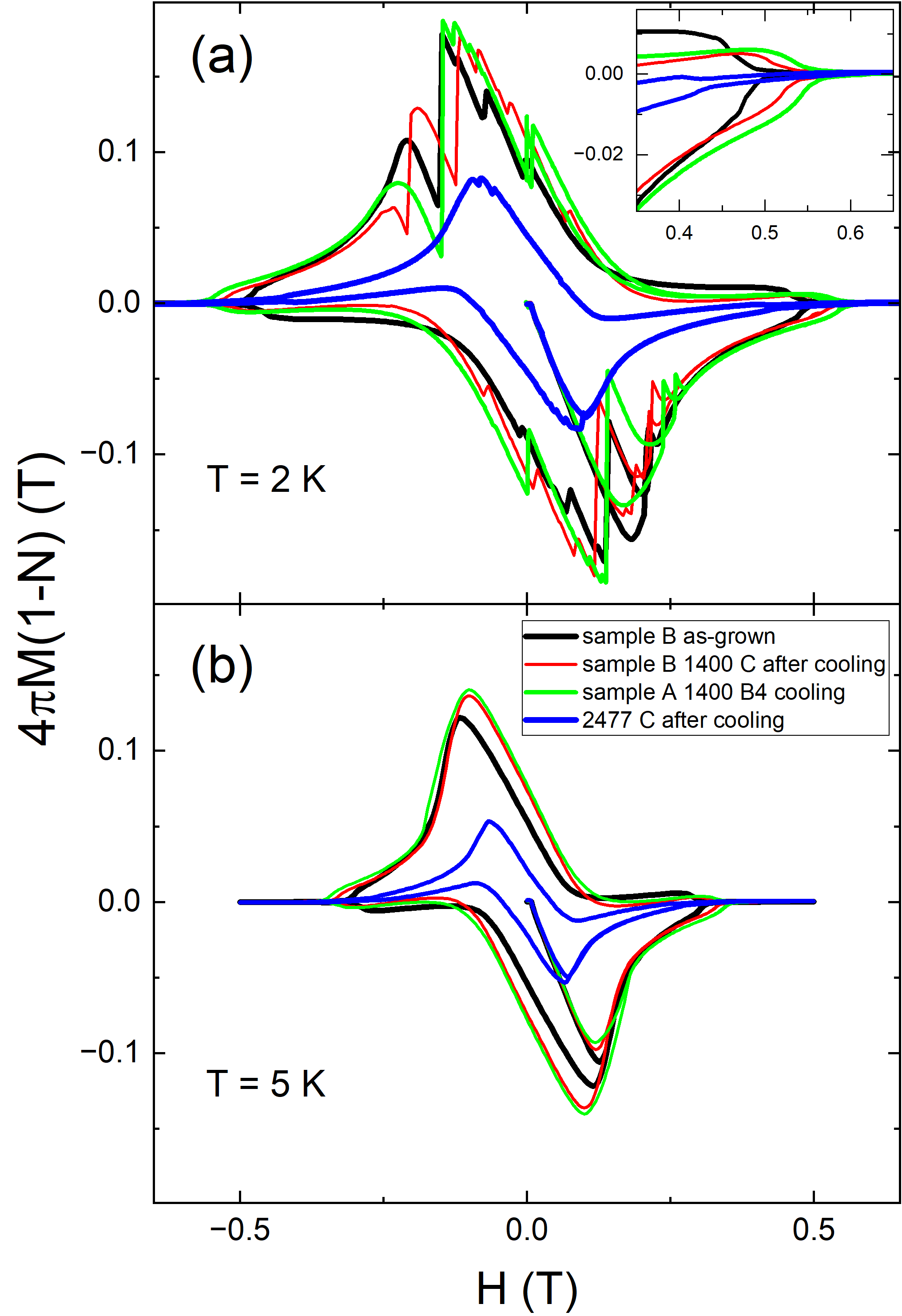} 
\caption{Comparison of $M(H)$ loops at (a) 2\:K and (b) 5\:K of the studied samples. The flux jump magnitude is the largest in sample A which was annealed before being cooled down. Inset in (a): detail of the higher field region showing that the upper critical field is the largest in the annealed sample A, followed by the annealed sample B and, finally, by the as-grown sample B confirming a commensurate sequence of non-magnetic scattering
strength.}
\label{fig7:M(T)5K2K}
\end{figure}

We further look at the magnitude of thermomagnetic instabilities in Fig.\ref{fig7:M(T)5K2K} comparing magnetization loops of all samples at 2\:K in panel (a) and at 5\:K in panel (b). The absence of flux jumps at 5:K confirms that their intrinsic nature is not related to any issues with sample structure or morphology. Clearly, sample A (bright green line) shows the largest avalanches indicative of stronger pinning. Sample B after annealing (red line) shows intermediate jump amplitudes compared to the smallest jumps in sample B before annealing (black line). The partially melted sample shows the smallest hysteresis and no thermomagnetic instabilities consistent with a complete removal of hydrogen, hence the sources of pinning. 

Another feature of Fig.\ref{fig7:M(T)5K2K} is the evolution of the upper critical field, $H_{c2}$. The inset in Fig.\ref{fig7:M(T)5K2K}(a) zooms in on the higher field region, showing that the upper critical field is the largest in annealed sample A, followed by annealed sample B, and finally by as-grown sample B. Previous studies found that $H_{c2}$ is insensitive to structural parameters such as tension and plastic deformation in crystals \cite{Tedmon1965}. In theory, $H_{c2}$ is linearly proportional to the non-magnetic scattering rate \cite{HW1966,p-irr-bridge} and this observation means that this rate increase after annealing points to the proliferation of nanosized defects that can act as efficient scattering centers. 

\begin{figure}[tb] 
\centering
\includegraphics[width=7cm]{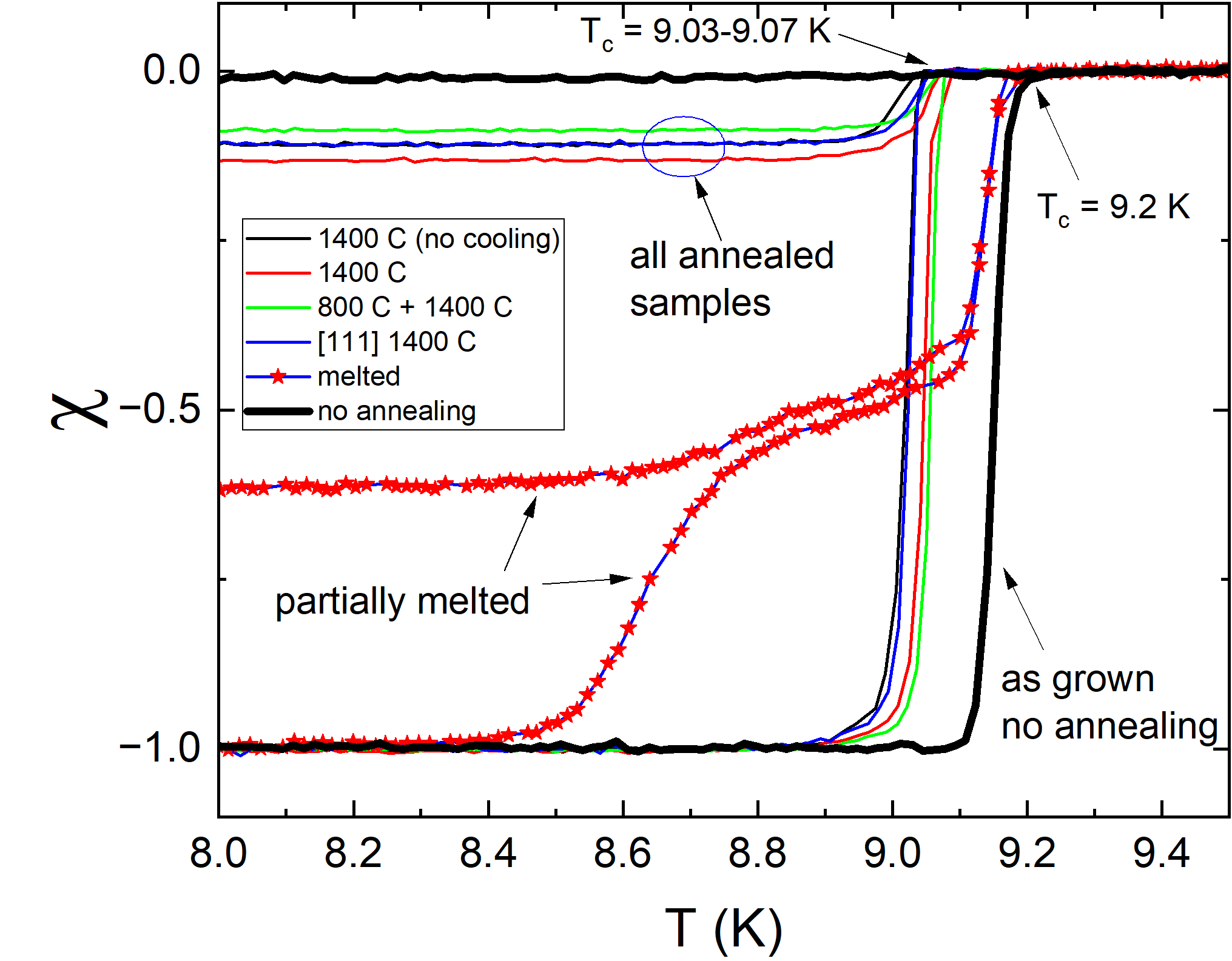} 
\caption{Comparison of the temperature dependent magnetic susceptibility, $\chi(T)$, of the studied samples. For each sample, both ZFC and FC traces are shown. Note that annealing reduces the transition temperature. Only the partially melted sample has recovered the original transition.}
\label{fig8:chi(T)}
\end{figure}

Another important parameter sensitive to the scattering rate is the superconducting transition temperature $T_{c}$. The anisotropy of the superconducting state of niobium has recently attracted significant attention, because of this anisotropy even non-magnetic scattering is pair-breaking, which is detrimental to the performance of QIS hardware \cite{Zarea2023}. Experimental studies confirm the theory \cite{p-irr-bridge}. 

Figure \ref{fig8:chi(T)} shows magnetic susceptibility as a function of the temperature recorded in the studied samples at 10\:Oe. Each sample was first cooled to 2\:K in zero field and a 10\:Oe magnetic field was applied. Then, the magnetization was measured on warming. After reaching temperatures above $T_{c}$, the measurements continued on cooling in the same magnetic field of 10\:Oe. Due to almost perfect shielding of a weak 10\:Oe magnetic field at 2\:K after ZFC, the magnetic susceptibility was set at -1 at that point, providing a natural calibration to the whole ZFC-FC curve.

As can be easily seen in Fig.\ref{fig8:chi(T)}, annealing reduces the transition temperature. Only the partially melted sample has recovered the original transition. At the same time, the initial curve of sample B measured before annealing shows the highest $T_{c}=9.2\;\mathrm{K}$. As we have already concluded, this indicates that large hydrides are not effective pair-breakers and do not lead to the reduction of $T_{c}$ seen. Instead, the evidence from magneto-optical and magnetization measurements for the enhanced pinning resulting from the increase of $H_{c2}$, which led to an increase of the non-magnetic scattering rate, together with a suppression of $T_{c}$, points to the proliferation of nanosized scattering/pinning centers as a result of annealing. The only candidates that we can think of are nano-hydrides. Of course, we cannot exclude some additional influence of oxygen ions that might have diffused into the bulk from the surface. However, the minuscule total amount of surface oxygen is not expected to have an impact on the bulk properties of the large single crystals studied here.

We also note that magnetic susceptibility values after field cooling to low temperatures are comparable for all samples, except for the partially melted sample. This reversible component is determined by the competition between Meissner expulsion and flux pinning, and this indicates that substantial changes could only be obtained after annealing at a very high temperature.

\subsection{Scanning tunneling microscopy (STM)}

\begin{figure}[tb] 
\centering
\includegraphics[width=7cm]{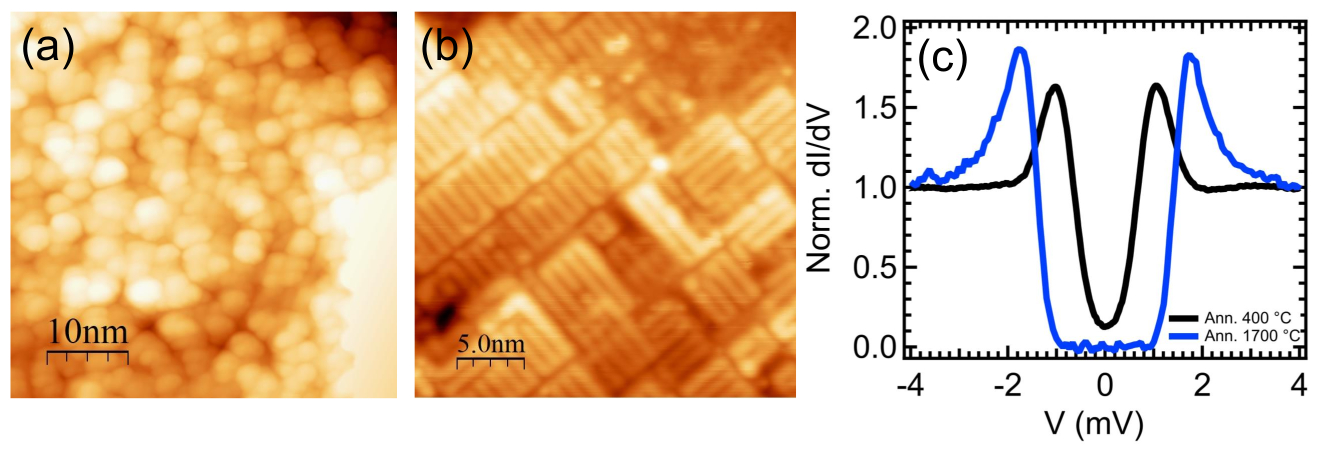} 
\caption{(a) STM topography image of a Nb (100) single crystal after annealing at 400\,\celsius. The topography has been acquired with tunneling conditions: $I=30\,\mathrm{pA}$, $V=1.0\,\mathrm{V}$. (b) Nb (100) topography after annealing at 1700\,\celsius\: clearly showing a $(3\times1)$ oxygen reconstruction. Tunneling conditions were: $I=30\,\mathrm{pA}$, $V=1.4\,\mathrm{V}$. (c) Typical tunneling
spectra acquired at $T=1.5\,\mathrm{K}$ on the two surfaces shown in panels (a) and (b). Tunneling conditions were $I=60\,\mathrm{pA}$, $V=10\,\mathrm{mV}$, and a lock-in modulation $V=0.2\,\mathrm{mV}$, and frequency $f=373.1\,\mathrm{Hz}$.}
\label{fig9:400C1700C}
\end{figure}

To examine the effect of annealing on the intrinsic superconducting properties and their (in)homogeneity at the atomic level, we conducted a scanning tunneling microscopy (STM) study. Importantly, the sample for the STM study was cut from the same master boule as the rest of the samples in this study. In the case of STM, surface condition plays a crucial role. To address this, STM samples were treated in separate procedures described in the experimental section. In addition to hydrogen, niobium has a high affinity for oxygen. Bulk oxygen impurities reduce the superconducting transition temperature of Nb at a rate of 1\:K per atomic percent \cite{Koch1974}. It is important to note that this study involved annealing in an oxygen-rich atmosphere, which is different from the present research where oxygen is only available in the surface oxide layers.

In the case of samples handled under normal laboratory conditions, the surface of Nb exposed to the atmosphere is always covered by a complex oxide layer of a few nanometers thick \cite{Kim2012,Murthy2022a}. An atom-probe study showed that oxygen diffuses to a depth of up to 10 nanometers from the surface, while hydrogen and niobium hydrides are present much deeper, essentially over the entire depth of their study of about 80 nm \cite{Kim2012}. Complete removal of niobium hydrides appears possible by annealing under ultra-high vacuum at temperatures higher than 2000\,\celsius\: and more (melting point of Nb is 2477\,\celsius) \cite{Tedmon1965,An_oxide_2003}. However, it is unknown whether surface oxygen can be completely removed without melting the sample. 

\begin{figure}[tb]
\centering
\includegraphics[width=7cm]{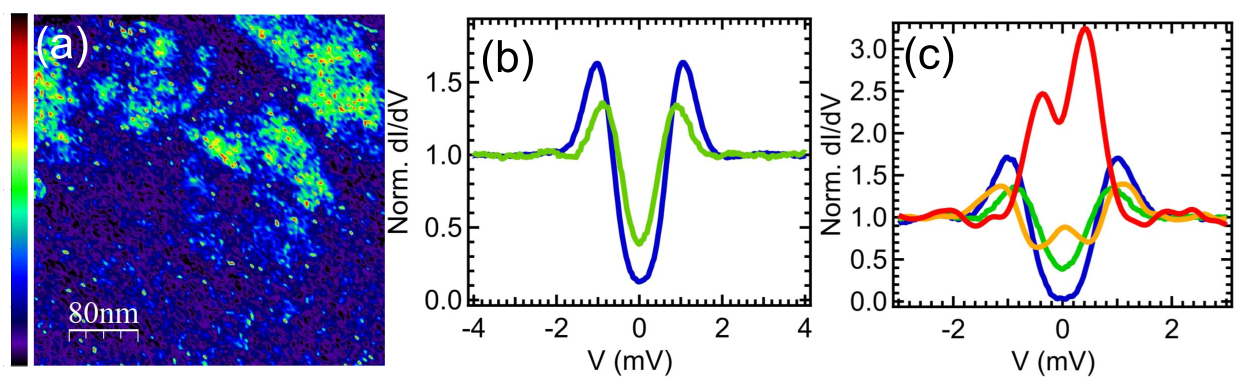} 
\caption{(a) Conductance map at the Fermi energy acquired on the surface of a Nb (100) single crystal after annealing at 400\,\celsius. (b) Tunneling spectra, $dI/dV$, as function of the bias voltage, $V$, acquired on the regions with the lowest conductance (blue) and on the regions with higher (but not the highest) conductance (green) on the same area shown in panel (a). (c) Tunneling spectra acquired at isolated regions on the map showing in-gap bound states (peaks at zero bias). Tunneling conditions for the conductance map and the spectra were $I=60\:\mathrm{pA}$, $V=10\:\mathrm{mV}$, and a lock-in modulation, $V=0.2\:\mathrm{mV}$ with a lock-in frequency of $373.1\:\mathrm{Hz}$.}
\label{fig10:STM-gap}
\end{figure}

In an initial experiment, the Nb (100) crystal was annealed at 400\,\celsius. Figure$\:\ref{fig9:400C1700C}$(a) shows a topography image of a scan area of $47\times47\:\mathrm{nm^{2}}$ showing a root mean square roughness of $1.5\:\mathrm{nm}$. At this stage, the complex surface oxide is expected to be metallic disordered NbO$_{x}$. In the next step, the same Nb (100) crystal was annealed at 1700\,\celsius. Figure$\:\ref{fig9:400C1700C}$(b) shows a $28\times28\:\mathrm{nm^{2}}$ topographic image of the reconstructed $(3\times1)$ oxygen structure of the NbO (100) layer on Nb (100) \cite{An_oxide_2003}. In Fig.$\,\ref{fig9:400C1700C}$(c), characteristic tunneling spectra acquired on these two surfaces at $T=1.5\:\mathrm{K}$ are shown. In the sample annealed at 400\,\celsius\: a significantly broadened tunneling spectrum is observed with a reduced gap of 1.35 meV (black curve). Although disordered oxygen ions near the surface occupied by a few nanometer thick NbO$_{x}$ certainly play a significant role \cite{Odobesko2019}, it is plausible that nanosized hydride precipitates further smear the tunneling curves. In the case of the sample annealed at 1700\,\celsius\: the conductance below the gap drops to zero and the superconducting gap is 1.55 meV (blue curve). No spatial variations in tunneling spectra were observed in the sample annealed at 1700\,\celsius, while the gap varied significantly on the surface of the same sample annealed at 400\,\celsius. It is quite possible that after this relatively low temperature annealing, intermediate-sized hydrides are formed, leading to a significant spatial variation when observed in a relatively small STM scan frame. However, oxygen redistribution near the surface must also play a significant role. A conductance map at the Fermi level acquired on this sample is shown in Fig.$\,\ref{fig10:STM-gap}$(a). Figure$\,\ref{fig10:STM-gap}$(b) shows the select conductance curves ranging from lower conductance (cleaner gap), shown by the blue curve, to areas of higher conductance, shown by the green curve. The map in Fig.$\,\ref{fig10:STM-gap}$(a) also shows nanometer-sized regions with much higher levels of conductance at Fermi energy, shown by the orange and red spectra in Fig.$\,\ref{fig10:STM-gap}$(c). These in-gap states, exhibiting conductance levels exceeding that of normal metals, suggest the presence of localized magnetic impurities\cite{YU1965,Shiba1968,Rusinov1969}. These impurities are likely associated with vacancies in the disordered NbO$_{x}$ top layer or in regions where the pentoxide has not been completely removed.

\section{Discussion}

We now discuss the results of the presented systematic study of the effect of annealing niobium single crystals with a significant initial hydrogen load. The high hydrogen content, a result of contact with water during normal sample cutting, polishing, etc. of the single crystal ingot, and storage under ambient conditions for many years, led to the formation of large hydride precipitates, of the order of 1-100\,{\micro m} and more when the samples were cooled. Although these hydrides significantly distort the crystal lattice, causing irreparable damage, they do not affect the microscopic superconducting properties of the surrounding niobium, such as pinning, critical temperature, and critical magnetic field. Of course, these large hydride formations significantly affect surface roughness, including some precipitates that rise above the surface by 6-8\,{\micro m} \cite{Barkov2013}, and surely degrade the performance of superconducting cavities \cite{Barkov2012,Barkov2013}. However, even moderate annealing at 800\,\celsius, importantly - performed before any cooling - prevents the formation of these large hydride precipitates, which seems to solve the problem. However, our findings point to the emergence of other potentially serious problems as a result of annealing.

Without microscopic spatially resolved elemental analysis, we can only suggest plausible scenarios. Regardless of the microscopic mechanisms, we have the following facts and conclusions. 

(1) Annealing at moderate temperatures, at least at 800\,\celsius, eliminates the formation of large precipitates of hydrides (tens and hundreds of micrometers). (2) Annealing at least up to 1400\,\celsius, leads to a reduction of the superconducting transition temperature, $T_c$, from its value in the same samples prior to annealing. Annealing very close to the melting point recovers the original $T_c$. (3) Annealing at 400\,\celsius\: shows inhomogeneous superconductivity with regions of dirty superconducting gap significantly contaminated by the in-gap states. In addition, there is direct evidence for zero-bias states in the gap consistent with magnetic impurities. This finding is consistent with the observation of a non-monotonic evolution of the quality factor of the Nb SRF resonators upon annealing \cite{Bafia2021}. Annealing at 1700\,\celsius\: recovers a clean superconducting gap. (4) The upper critical field increases after annealing at intermediate temperatures. The magnitude of the effect is correlated with the magnitude of suppression of $T_c$. (5) Contrary to expectations of decrease, the pinning strength increases moderately after annealing. (6) Only partial melting leads to a significant reduction in pinning, a much larger Meissner effect, and the recovery of $T_c$.

A large portion of the hydrogen absorbed by Nb from its environment can be removed by relatively low-temperature annealing. However, we suggest that some dilute hydrogen subsystem, bound or not, remains, and upon cooling it leads to the formation of nanosized niobium hydride precipitates. Together, our results indicate that annealing leads to the proliferation of defects that affect superconductivity, so they must be of the order of the coherence length, 10-40\,{nm}. Due to the gap anisotropy, those nonmagnetic scattering centers suppress $T_c$. At the same time, the upper critical field and the pinning strength increase. 

Another potential scenario is that during annealing, oxygen from the surface oxide layers diffuses into the bulk. However, it was determined that niobium oxides are mainly distributed within the first 10 micrometers of the surface \cite{Halbritter1988}. If the suppression of the $T_c$ we observed was due to such oxygen ions, the transition curves would not parallel shift, but would become smeared, which was not the case here. The same can be said with respect to the upper critical field, which is determined by the entire vortex length of 1-2\:mm in our case, and the small amount of oxygen from a surface layer would not change its bulk values. Since it has been shown that high temperature vacuum annealing at 2300\,\celsius\: for several hours significantly reduces background oxygen \cite{Tedmon1965,Halbritter1988}, it can be minimized by a pre-annealing step if needed. Finally, as estimated in the Introduction, the total amount of oxygen in the oxide layers is way too small to cause a noticeable variation of bulk properties.

We stress that apart from a pronounced effect of the suppression of macroscopic hydrides, the changes upon annealing discussed in this paper are subtle. However, these issues are potentially important for delicate QIS applications and, perhaps mitigating them holds the key to a substantial improvement of quantum hardware.

\section{Conclusions}

A systematic study of the effects of annealing on the superconducting properties of niobium crystals was carried out using optical, magnetic, and tunneling spectroscopy measurements. It is shown that fairly accessible annealing at 800\,\celsius, indeed, suppresses the formation of very large hydrides, probably due to the removal of most of the absorbed hydrogen. However, such annealing does not eliminate all hydrogen and introduces new problems associated with impurities that affect superconductivity. It appears that high-vacuum annealing at temperatures, perhaps exceeding 1700\,\celsius, is needed to obtain niobium for applications that require ultra-low losses suitable for QIS applications. We note that many studies of the 1960-1980s routinely used UHV annealing above 2000\,\celsius\: for a few hours to obtain high-quality samples free of not only mechanical deformation, but also impurities such as oxygen and hydrogen \cite{Tedmon1965,Halbritter1988}.

\section{Acknowledgements}

We thank Arne Swanson of the Materials Processing Center at Ames National
Laboratory for running the 1400$\:\celsius$ annealing. We thank Thomas
Lograsso for growing the original Nb crystal. This work was supported
by the U.S. Department of Energy, Office of Science, National Quantum
Information Science Research Centers, Superconducting Quantum Materials
and Systems Center (SQMS) under contract number DE-AC02-07CH11359.
The research was carried out at the Ames Laboratory, operated for the
U.S. DOE by Iowa State University under contract \# DE-AC02-07CH11358.

 \bibliographystyle{elsarticle-num} 

\begin{thebibliography}{10}
\expandafter\ifx\csname url\endcsname\relax
  \def\url#1{\texttt{#1}}\fi
\expandafter\ifx\csname urlprefix\endcsname\relax\def\urlprefix{URL }\fi
\expandafter\ifx\csname href\endcsname\relax
  \def\href#1#2{#2} \def\path#1{#1}\fi

\bibitem{Grezes2016}
C.~Gr{\`{e}}zes, Towards a Spin-Ensemble Quantum Memory for Superconducting
  Qubits, 1st Edition, Springer, 2016.

\bibitem{Kjaergaard2020}
M.~Kjaergaard, M.~E. Schwartz, J.~Braum\"{u}ller, P.~Krantz, J.~I.-J. Wang,
  S.~Gustavsson, W.~D. Oliver, Superconducting qubits: Current state of play,
  Annu. Rev. Conden. Ma. P. 11~(1) (2020) 369--395.
\newblock \href {https://doi.org/10.1146/annurev-conmatphys-031119-050605}
  {\path{doi:10.1146/annurev-conmatphys-031119-050605}}.

\bibitem{Huang2020}
H.-L. Huang, D.~Wu, D.~Fan, X.~Zhu, Superconducting quantum computing: a
  review, Sci. China. Inform. Sci. 63~(8) (Jul. 2020).
\newblock \href {https://doi.org/10.1007/s11432-020-2881-9}
  {\path{doi:10.1007/s11432-020-2881-9}}.

\bibitem{He2021}
K.~He, X.~Geng, R.~Huang, J.~Liu, W.~Chen, Quantum computation and simulation
  with superconducting qubits{\textasteriskcentered}, Chinese Phys. B 30~(8)
  (2021) 080304.
\newblock \href {https://doi.org/10.1088/1674-1056/ac16cf}
  {\path{doi:10.1088/1674-1056/ac16cf}}.

\bibitem{Siddiqi2021}
I.~Siddiqi, Engineering high-coherence superconducting qubits, Nat. Rev. Mater.
  6~(10) (2021) 875--891.
\newblock \href {https://doi.org/10.1038/s41578-021-00370-4}
  {\path{doi:10.1038/s41578-021-00370-4}}.

\bibitem{Xiong2022}
K.~Xiong, J.~Feng, Y.~Zheng, J.~Cui, M.~Yung, S.~Zhang, S.~Li, H.~Yang,
  Materials in superconducting quantum circuits, Chinese Science
  Bulletin-chinese 67~(2) (2022) 143--162.
\newblock \href {https://doi.org/10.1360/tb-2021-0479}
  {\path{doi:10.1360/tb-2021-0479}}.

\bibitem{Ezratty2023}
O.~Ezratty, Perspective on superconducting qubit quantum computing, Eur. Phys.
  J. A 59~(5) (2023).
\newblock \href {https://doi.org/10.1140/epja/s10050-023-01006-7}
  {\path{doi:10.1140/epja/s10050-023-01006-7}}.

\bibitem{padamsee2009}
H.~Padamsee, RF Superconductivity, 1st Edition, Wiley-VCH, Weinheim, 2009.

\bibitem{Chao2014}
A.~W. Chao, Applications of superconducting technology to accelerators, in:
  W.~Chou (Ed.), Reviews of Accelerator Science and Technology, no.~5 in
  Reviews of accelerator science and technology, World Scientific Publishing Co
  Pte Ltd, Singapore, 2014, p. 1369.

\bibitem{Anne_2022}
A.~M. Valente-Feliciano, C.~Antoine, S.~Anlage, G.~Ciovati, J.~Delayen,
  F.~Gerigk, A.~Gurevich, T.~Junginger, S.~Keckert, G.~Keppe, J.~Knobloch,
  T.~Kubo, O.~Kugeler, D.~Manos, C.~Pira, T.~Proslier, U.~Pudasaini, C.~E.
  Reece, R.~A. Rimmer, G.~J. Rosaz, T.~Saeki, R.~Vaglio, R.~Valizadeh,
  H.~Vennekate, W.~V. Delsolaro, M.~Vogel, P.~B. Welander, M.~Wenskat,
  Next-generation superconducting rf technology based on advanced thin film
  technologies and innovative materials for accelerator enhanced performance
  and energy reach, arXiv:2204.02536 (2022).

\bibitem{Stromberg1965}
T.~F. Stromberg, The superconducting properties of high purity niobium, Ph.D.
  thesis, Iowa State University (1965).

\bibitem{Finnemore1966}
D.~K. Finnemore, T.~F. Stromberg, C.~A. Swenson, Superconducting properties of
  high-purity niobium, Phys. Rev. 149~(1) (1966) 231--243.
\newblock \href {https://doi.org/10.1103/physrev.149.231}
  {\path{doi:10.1103/physrev.149.231}}.

\bibitem{Daams1980}
J.~Daams, J.~P. Carbotte, {Thermodynamic properties of superconducting
  niobium}, J. Low Temp. Phys. 40~(1) (1980) 135--154.
\newblock \href {https://doi.org/10.1007/bf00115987}
  {\path{doi:10.1007/bf00115987}}.

\bibitem{Bahte1998}
M.~Bahte, F.~Herrmann, P.~Schmuser, {Magnetization and susceptibility
  measurements on niobium samples for cavity production [8th Workshop on RF
  Superconductivity]}, Part. Accel. 60~(1-4) (1998) 121--133.

\bibitem{Koethe2000}
A.~Koethe, J.~I. Moench, Preparation of ultra high purity niobium, Mater.
  Trans., JIM 41~(1) (2000) 7--16.

\bibitem{Prozorov2006a}
R.~Prozorov, D.~V. Shantsev, R.~G. Mints, Collapse of the critical state in
  superconducting niobium, Phys. Rev. B 74 (2006) 220511.
\newblock \href {https://doi.org/10.1103/PhysRevB.74.220511}
  {\path{doi:10.1103/PhysRevB.74.220511}}.

\bibitem{Kozhevnikov2017}
V.~Kozhevnikov, A.-M. Valente-Feliciano, P.~J. Curran, A.~Suter, A.~H. Liu,
  G.~Richter, E.~Morenzoni, S.~J. Bending, C.~V. Haesendonck, Equilibrium
  properties of superconducting niobium at high magnetic fields: A possible
  existence of a filamentary state in type-{II} superconductors, Phys. Rev. B
  95~(17) (2017) 174509.
\newblock \href {https://doi.org/10.1103/physrevb.95.174509}
  {\path{doi:10.1103/physrevb.95.174509}}.

\bibitem{Liarte2017}
D.~B. Liarte, S.~Posen, M.~K. Transtrum, G.~Catelani, M.~Liepe, J.~P. Sethna,
  Theoretical estimates of maximum fields in superconducting resonant radio
  frequency cavities: stability theory, disorder, and laminates, Supercond.
  Sci. Technol. 30~(3) (2017) 033002.
\newblock \href {https://doi.org/10.1088/1361-6668/30/3/033002}
  {\path{doi:10.1088/1361-6668/30/3/033002}}.

\bibitem{wendin2017quantum}
G.~Wendin, Quantum information processing with superconducting circuits: a
  review, Rep. Prog. Phys. 80~(10) (2017) 106001.

\bibitem{Padamsee2008}
H.~Padamsee, J.~Knobloch, T.~Hays, RF Superconductivity for Accelerators,
  Wiley, 2008.

\bibitem{Gurevich2012}
A.~Gurevich, Superconducting radio-frequency fundamentals for particle
  accelerators, Rev. Accel Sci. Technol. 05 (2012) 119--146.
\newblock \href {https://doi.org/10.1142/s1793626812300058}
  {\path{doi:10.1142/s1793626812300058}}.

\bibitem{Singer2014}
{Singer, W}, {SRF Cavity Fabrication and Materials}, {arXiv:1501.07142}
  ({2014}).
\newblock \href {https://doi.org/{10.5170/CERN-2014-005.171}}
  {\path{doi:{10.5170/CERN-2014-005.171}}}.

\bibitem{Gurevich2017a}
A.~Gurevich, Theory of {RF} superconductivity for resonant cavities, Supercond.
  Sci. Technol. 30~(3) (2017) 034004.
\newblock \href {https://doi.org/10.1088/1361-6668/30/3/034004}
  {\path{doi:10.1088/1361-6668/30/3/034004}}.

\bibitem{alex20}
A.~Romanenko, R.~Pilipenko, S.~Zorzetti, D.~Frolov, M.~Awida, S.~Belomestnykh,
  S.~Posen, A.~Grassellino, {Three-Dimensional Superconducting Resonators at
  T=20 mK with Photon Lifetimes up to $\tau=2$\,s}, Phys. Rev. Appl. 13 (2020)
  034032.
\newblock \href {https://doi.org/10.1103/PhysRevApplied.13.034032}
  {\path{doi:10.1103/PhysRevApplied.13.034032}}.

\bibitem{Ueki2022}
H.~Ueki, M.~Zarea, J.~A. Sauls, The frequency shift and q of disordered
  superconducting rf cavities, arXiv:2207.14236 (2022).

\bibitem{Ueki2022a}
H.~Ueki, M.~Zarea, J.~A. Sauls, Electromagnetic response of superconducting rf
  cavities, arXiv:2209.11752 (2022).

\bibitem{Blais2004}
A.~Blais, R.-S. Huang, A.~Wallraff, S.~M. Girvin, R.~J. Schoelkopf, Cavity
  quantum electrodynamics for superconducting electrical circuits: An
  architecture for quantum computation, Phys. Rev. A 69~(6) (2004) 062320.
\newblock \href {https://doi.org/10.1103/physreva.69.062320}
  {\path{doi:10.1103/physreva.69.062320}}.

\bibitem{Paik2011}
H.~Paik, D.~I. Schuster, L.~S. Bishop, G.~Kirchmair, G.~Catelani, A.~P. Sears,
  B.~R. Johnson, M.~J. Reagor, L.~Frunzio, L.~I. Glazman, S.~M. Girvin, M.~H.
  Devoret, R.~J. Schoelkopf, Observation of high coherence in josephson
  junction qubits measured in a three-dimensional circuit {QED} architecture,
  Phys. Rev. Lett. 107~(24) (2011) 240501.
\newblock \href {https://doi.org/10.1103/physrevlett.107.240501}
  {\path{doi:10.1103/physrevlett.107.240501}}.

\bibitem{Reagor2016}
M.~Reagor, W.~Pfaff, C.~Axline, R.~W. Heeres, N.~Ofek, K.~Sliwa, E.~Holland,
  C.~Wang, J.~Blumoff, K.~Chou, M.~J. Hatridge, L.~Frunzio, M.~H. Devoret,
  L.~Jiang, R.~J. Schoelkopf, Quantum memory with millisecond coherence in
  circuit {QED}, Phys. Rev. B 94~(1) (2016) 014506.
\newblock \href {https://doi.org/10.1103/physrevb.94.014506}
  {\path{doi:10.1103/physrevb.94.014506}}.

\bibitem{Daunt1937}
J.~G. Daunt, K.~Mendelssohn, F.~A. Lindemann, Equilibrium curve and entropy
  difference between the supraconductive and the normal state in pb, hg, sn,
  ta, and nb, Proc. Roy. Soc. London. Ser. A - Math. Phys. Sci. 160~(900)
  (1937) 127--136.
\newblock \href {https://doi.org/10.1098/rspa.1937.0099}
  {\path{doi:10.1098/rspa.1937.0099}}.

\bibitem{Kneisel2015}
P.~Kneisel, G.~Ciovati, P.~Dhakal, K.~Saito, W.~Singer, X.~Singer, G.~R.
  Myneni, Review of ingot niobium as a material for superconducting
  radiofrequency accelerating cavities, Nucl. Instrum. Methods Phys. Res.,
  Sect. A 774 (2015) 133--150.
\newblock \href {https://doi.org/10.1016/j.nima.2014.11.083}
  {\path{doi:10.1016/j.nima.2014.11.083}}.

\bibitem{Zarea2023}
M.~Zarea, H.~Ueki, J.~A. Sauls, Effects of anisotropy and disorder on the
  superconducting properties of niobium, Aip. Conf. Proc. 11 (2023) 1269872.
\newblock \href {https://doi.org/10.3389/fphy.2023.1269872}
  {\path{doi:10.3389/fphy.2023.1269872}}.

\bibitem{Prozorov2022}
R.~Prozorov, M.~Zarea, J.~A. Sauls, {Niobium in the clean limit: An intrinsic
  type-I superconductor}, Phys. Rev. B 106~(18) (2022) L180505.
\newblock \href {https://doi.org/10.1103/PhysRevB.106.L180505}
  {\path{doi:10.1103/PhysRevB.106.L180505}}.

\bibitem{Knobloch2003}
J.~Knobloch, {The ``Q disease'' in Superconducting Niobium RF Cavities}, AIP
  Conf. Proc. 671~(1) (2003) 133--150.
\newblock \href {https://doi.org/10.1063/1.1597364}
  {\path{doi:10.1063/1.1597364}}.

\bibitem{Barkov2012}
F.~Barkov, A.~Romanenko, A.~Grassellino, Direct observation of hydrides
  formation in cavity-grade niobium, Phys. Rev. Spec. Top. Accel Beams 15~(12)
  (2012) 122001.
\newblock \href {https://doi.org/10.1103/physrevstab.15.122001}
  {\path{doi:10.1103/physrevstab.15.122001}}.

\bibitem{Barkov2013}
F.~Barkov, A.~Romanenko, Y.~Trenikhina, A.~Grassellino, Precipitation of
  hydrides in high purity niobium after different treatments, J. Appl. Phys.
  114~(16) (2013) 164904.
\newblock \href {https://doi.org/10.1063/1.4826901}
  {\path{doi:10.1063/1.4826901}}.

\bibitem{Buck1971}
O.~Buck, D.~Thompson, C.~Wert, The effect of hydrogen on the low temperature
  internal friction of a niobium single crystal, J. Phys. Chem. Sol. 32~(10)
  (1971) 2331--2344.
\newblock \href {https://doi.org/10.1016/s0022-3697(71)80227-5}
  {\path{doi:10.1016/s0022-3697(71)80227-5}}.

\bibitem{Schober1975}
T.~Schober, The niobium–hydrogen system – an electron microscope study. ii.
  low-temperature structures, phys. stat. sol. (a) 30~(1) (1975) 107--116.
\newblock \href {https://doi.org/10.1002/pssa.2210300111}
  {\path{doi:10.1002/pssa.2210300111}}.

\bibitem{Isagawa1980}
S.~Isagawa, Influence of hydrogen on superconducting niobium cavities, J. Appl.
  Phys. 51~(11) (1980) 6010--6017.
\newblock \href {https://doi.org/10.1063/1.327523}
  {\path{doi:10.1063/1.327523}}.

\bibitem{Isagawa1980a}
S.~Isagawa, Hydrogen absorption and its effect on low-temperature electric
  properties of niobium, Journal of Applied Physics 51~(8) (1980) 4460--4470.
\newblock \href {https://doi.org/10.1063/1.328267}
  {\path{doi:10.1063/1.328267}}.

\bibitem{Dzyuba2014}
A.~Dzyuba, L.~D. Cooley, Combined effects of cold work and chemical polishing
  on the absorption and release of hydrogen from srf cavities inferred from
  resistance measurements of cavity-grade niobium bars, Superconductor Science
  and Technology 27~(3) (2014) 035001.
\newblock \href {https://doi.org/10.1088/0953-2048/27/3/035001}
  {\path{doi:10.1088/0953-2048/27/3/035001}}.

\bibitem{Vinnikov1982}
L.~Y. Vinnikov, A.~O. Golubok, Direct observation of magnetic structure in
  niobium single crystals with hydride precipitate pinning centres, Phys. Stat.
  Sol. A 69~(2) (1982) 631--636.
\newblock \href {https://doi.org/10.1002/pssa.2210690224}
  {\path{doi:10.1002/pssa.2210690224}}.

\bibitem{Koszegi2017}
J.~K\"{o}szegi, O.~Kugeler, D.~Abou-Ras, J.~Knobloch, R.~Sch\"{a}fer, A
  magneto-optical study on magnetic flux expulsion and pinning in high-purity
  niobium, J. Appl. Phys. 122~(17) (2017) 173901.
\newblock \href {https://doi.org/10.1063/1.4996113}
  {\path{doi:10.1063/1.4996113}}.

\bibitem{Burnett2016}
J.~Burnett, L.~Faoro, T.~Lindstrom, Analysis of high quality superconducting
  resonators: consequences for {TLS} properties in amorphous oxides, Supercond.
  Sci. Technol. 29~(4) (2016) 044008.
\newblock \href {https://doi.org/10.1088/0953-2048/29/4/044008}
  {\path{doi:10.1088/0953-2048/29/4/044008}}.

\bibitem{Mueller2019}
C.~M\"{u}ller, J.~H. Cole, J.~Lisenfeld, Towards understanding
  two-level-systems in amorphous solids: insights from quantum circuits, Rep.
  Prog. Phys. 82~(12) (2019) 124501.
\newblock \href {https://doi.org/10.1088/1361-6633/ab3a7e}
  {\path{doi:10.1088/1361-6633/ab3a7e}}.

\bibitem{McRae20}
C.~R.~H. McRae, H.~Wang, J.~Gao, M.~R. Vissers, T.~Brecht, A.~Dunsworth, D.~P.
  Pappas, J.~Mutus, {Materials loss measurements using superconducting
  microwave resonators}, Rev. Sci. Instrum. 91~(9) (2020) 091101.
\newblock \href {https://doi.org/10.1063/5.0017378}
  {\path{doi:10.1063/5.0017378}}.

\bibitem{Tedmon1965}
C.~S. Tedmon, R.~M. Rose, J.~Wulff, Resistive measurements of structural
  effects in superconducting niobium, J. Appl.Phys. 36~(3) (1965) 829--835.
\newblock \href {https://doi.org/10.1063/1.1714227}
  {\path{doi:10.1063/1.1714227}}.

\bibitem{Halbritter1988}
J.~Halbritter, Low temperature oxidation of nb and of nb-compounds in relation
  to superconducting application, Journal of the Less Common Metals 139~(1)
  (1988) 133--148.
\newblock \href {https://doi.org/10.1016/0022-5088(88)90336-0}
  {\path{doi:10.1016/0022-5088(88)90336-0}}.

\bibitem{Antoine2019}
C.~Z. Antoine, Influence of crystalline structure on rf dissipation in
  superconducting niobium, Phys. Rev. Accel. Beams 22~(3) (2019) 034801.
\newblock \href {https://doi.org/10.1103/physrevaccelbeams.22.034801}
  {\path{doi:10.1103/physrevaccelbeams.22.034801}}.

\bibitem{DangwalPandey2021}
A.~Dangwal~Pandey, T.~F. Keller, M.~Wenskat, A.~Jeromin, S.~Kulkarni, H.~Noei,
  V.~Vonk, W.~Hillert, D.~Reschke, N.~Walker, H.~Weise, A.~Stierle, Grain
  boundary segregation and carbide precipitation in heat treated niobium
  superconducting radio frequency cavities, Appl. Phys. Lett. 119~(19) (2021)
  194102.
\newblock \href {https://doi.org/10.1063/5.0063379}
  {\path{doi:10.1063/5.0063379}}.

\bibitem{Hakovirta2001}
M.~Hakovirta, {Measurements of hydrogen content in bulk niobium by Thermal
  Desorption Spectroscopy}, CERN-EST-2002-005-SM,
  https://cds.cern.ch/record/572772 (2001).

\bibitem{Ooi2021}
S.~Ooi, M.~Tachiki, T.~Konomi, T.~Kubo, A.~Kikuchi, S.~Arisawa, H.~Ito,
  K.~Umemori, Observation of intermediate mixed state in high-purity
  cavity-grade nb by magneto-optical imaging, Phys. Rev. B 104~(6) (2021)
  064504.
\newblock \href {https://doi.org/10.1103/physrevb.104.064504}
  {\path{doi:10.1103/physrevb.104.064504}}.

\bibitem{Lograsso1991}
T.~A. Lograsso, F.~A. Schmidt, Solid-state crystal growth of refractory metals
  by arc-zone melting, J. Cryst. Growth 110~(3) (1991) 363--372.
\newblock \href {https://doi.org/10.1016/0022-0248(91)90272-7}
  {\path{doi:10.1016/0022-0248(91)90272-7}}.

\bibitem{King1990}
W.~E. King, G.~H. Campbell, A.~Coombs, M.~J. Mills, M.~R\"{u}hle, Hrem
  investigation of the structure of the \ensuremath{\sigma}5(310)/[001]
  symmetric tilt grain boundary in nb, MRS Proc. 209 (1990).
\newblock \href {https://doi.org/10.1557/proc-209-39}
  {\path{doi:10.1557/proc-209-39}}.

\bibitem{Prozorov2018}
R.~Prozorov, V.~G. Kogan, Effective demagnetizing factors of diamagnetic
  samples of various shapes, Phys. Rev. Applied 10 (2018) 014030.
\newblock \href {https://doi.org/10.1103/PhysRevApplied.10.014030}
  {\path{doi:10.1103/PhysRevApplied.10.014030}}.

\bibitem{Young2005}
D.~P. Young, M.~Moldovan, P.~W. Adams, R.~Prozorov, {M}agneto-optical studies
  of flux penetratim in super-hard {N}b wire, Superc. Sci. Technol. 18 (2005)
  776--779.

\bibitem{Joshi2022a}
K.~R. Joshi, S.~Ghimire, M.~A. Tanatar, A.~Datta, J.-S. Oh, L.~Zhou, C.~J.
  Kopas, J.~Marshall, J.~Y. Mutus, J.~Slaughter, M.~J. Kramer, J.~A. Sauls,
  R.~Prozorov, {Quasiparticle Spectroscopy, Transport, and Magnetic Properties
  of Nb Films Used in Superconducting Qubits}, Phys. Rev. Appl. 20~(2) (2023)
  024031.
\newblock \href {https://doi.org/10.1103/PhysRevApplied.20.024031}
  {\path{doi:10.1103/PhysRevApplied.20.024031}}.

\bibitem{HW1966}
E.~Helfand, N.~R. Werthamer, Temperature and purity dependence of the
  superconducting critical field, ${H}_{c2} .$ ii, Phys. Rev. 147 (1966)
  288--294.
\newblock \href {https://doi.org/10.1103/PhysRev.147.288}
  {\path{doi:10.1103/PhysRev.147.288}}.

\bibitem{p-irr-bridge}
M.~A. Tanatar, D.~Torsello, K.~R. Joshi, S.~Ghimire, C.~J. Kopas, J.~Marshall,
  J.~Y. Mutus, G.~Ghigo, M.~Zarea, J.~A. Sauls, R.~Prozorov, Anisotropic
  superconductivity of niobium based on its response to nonmagnetic disorder,
  Phys. Rev. B 106 (2022) 224511.
\newblock \href {https://doi.org/10.1103/PhysRevB.106.224511}
  {\path{doi:10.1103/PhysRevB.106.224511}}.

\bibitem{Koch1974}
C.~C. Koch, J.~O. Scarbrough, D.~M. Kroeger, Effects of interstitial oxygen on
  the superconductivity of niobium, Phys. Rev. B 9~(3) (1974) 888--897.
\newblock \href {https://doi.org/10.1103/physrevb.9.888}
  {\path{doi:10.1103/physrevb.9.888}}.

\bibitem{Kim2012}
Y.-J. Kim, R.~Tao, R.~F. Klie, D.~N. Seidman, Direct atomic-scale imaging of
  hydrogen and oxygen interstitials in pure niobium using atom-probe tomography
  and aberration-corrected scanning transmission electron microscopy, ACS Nano
  7~(1) (2012) 732--739.
\newblock \href {https://doi.org/10.1021/nn305029b}
  {\path{doi:10.1021/nn305029b}}.

\bibitem{Murthy2022a}
A.~A. Murthy, P.~Masih~Das, S.~M. Ribet, C.~Kopas, J.~Lee, M.~J. Reagor,
  L.~Zhou, M.~J. Kramer, M.~C. Hersam, M.~Checchin, A.~Grassellino, R.~d. Reis,
  V.~P. Dravid, A.~Romanenko, Developing a chemical and structural
  understanding of the surface oxide in a niobium superconducting qubit, ACS
  Nano 16~(10) (2022) 17257--17262.
\newblock \href {https://doi.org/10.1021/acsnano.2c07913}
  {\path{doi:10.1021/acsnano.2c07913}}.

\bibitem{An_oxide_2003}
B.~An, S.~Fukuyama, K.~Yokogawa, M.~Yoshimura, Surface structures of clean and
  oxidized nb(100) by leed, aes, and stm, Phys. Rev. B 68 (2003) 115423.
\newblock \href {https://doi.org/10.1103/PhysRevB.68.115423}
  {\path{doi:10.1103/PhysRevB.68.115423}}.

\bibitem{Odobesko2019}
A.~B. Odobesko, S.~Haldar, S.~Wilfert, J.~Hagen, J.~Jung, N.~Schmidt, P.~Sessi,
  M.~Vogt, S.~Heinze, M.~Bode, Preparation and electronic properties of clean
  superconducting nb(110) surfaces, Phys. Rev. B 99~(11) (2019) 115437.
\newblock \href {https://doi.org/10.1103/physrevb.99.115437}
  {\path{doi:10.1103/physrevb.99.115437}}.

\bibitem{YU1965}
L.~Yu, Bound state in superconductors with paramagnetic impurities, Acta Phys.
  Sin-ch. Ed. 21~(1) (1965) 75.
\newblock \href {https://doi.org/10.7498/aps.21.75}
  {\path{doi:10.7498/aps.21.75}}.

\bibitem{Shiba1968}
H.~Shiba, Classical spins in superconductors, Prog. Theor. Phys. 40~(3) (1968)
  435--451.
\newblock \href {https://doi.org/10.1143/ptp.40.435}
  {\path{doi:10.1143/ptp.40.435}}.

\bibitem{Rusinov1969}
A.~I. Rusinov, On the theory of gapless superconductivity in alloys containing
  paramagnetic impurities, [Zh. Eksp. Teor. Fiz. 56, 2047-2056 (1969)] Sov.
  Phys. JETP (1969) 1101--1106.

\bibitem{Bafia2021}
D.~Bafia, A.~Grassellino, A.~Romanenko, {Probing the Role of Low Temperature
  Vacuum Baking on Photon Lifetimes in Superconducting Niobium 3-D Resonators},
  arXiv:2108.13352 (2021).

\end{thebibliography}

\end{document}